\documentclass[runningheads]{llncs}
\usepackage[T1]{fontenc}
\usepackage{graphicx}
\usepackage{amsmath,amssymb}
\usepackage{booktabs}
\usepackage{multirow}
\usepackage{array}
\usepackage{xcolor}
\usepackage{colortbl}
\usepackage{tabularx}
\usepackage{caption}
\usepackage{pifont}
\usepackage[colorlinks,linkcolor=blue,citecolor=blue]{hyperref}
\usepackage[table,xcdraw]{xcolor}

\begin{document}

\title{FermatSyn: SAM2-Enhanced Bidirectional Mamba with Isotropic Spiral Scanning for Multi-Modal Medical Image Synthesis}
\titlerunning{FermatSyn: Fermat Spiral Mamba for Medical Image Synthesis}
\author{Feng Yuan\inst{1,3}, 
Yifan Gao\inst{1,2,3},  
Haoyue Li\inst{1,3}, 
Xin Gao\inst{3}\textsuperscript{\ding{41}} } 
\authorrunning{Feng Yuan et al.}
\institute{
    School of Biomedical Engineering (Suzhou), Division of Life Science and Medicine, University of Science and Technology of China, Hefei, China \and
    Shanghai Innovation Institute, Shanghai, China \and
    Suzhou Institute of Biomedical Engineering and Technology, Chinese Academy of Sciences, Suzhou, China \\
    \email{yuanfeng2317@mail.ustc.edu.cn} \\
    \email{xingaosam@163.com}
}
\maketitle              

\begin{abstract}
Multi-modal medical image synthesis is pivotal for alleviating
clinical data scarcity, yet existing methods fail to reconcile
global anatomical consistency with high-fidelity local detail.
We propose FermatSyn, which addresses three persistent
limitations:
(1) SAM2-based Prior Encoder that injects domain-aware
anatomical knowledge via LoRA$^{+}$ efficient fine-tuning of a
frozen SAM2 Vision Transformer;
(2) Hierarchical Residual Downsampling Module (HRDM)
coupled with a Cross-scale Integration Network (CIN) that
preserves high-frequency lesion details and adaptively fuses
global--local representations; and
(3) continuity constrained Fermat Spiral Scanning
strategy within a Bidirectional Fermat Scan Mamba (BFS-Mamba),
constructing an approximately isotropic receptive field that
substantially reduces the directional bias of raster or spiral serialization.
Experiments on SynthRAD2023, BraTS2019, BraTS-MEN, and BraTS-MET
show FermatSyn surpasses state-of-the-art methods in PSNR, SSIM,
FID, and 3D structural consistency.
Downstream segmentation on synthesized images yields no significant
difference from real-image training ($p{>}0.05$), confirming
clinical utility. Code is available at \url{https://github.com/exsinger-hub/FermatSyn}.

\end{abstract}

\section{Introduction}
\label{sec:intro}

Acquiring complete multi-modal medical imaging datasets is routinely
impractical owing to prolonged scan times, patient
contraindications~\cite{thukral2015problems}, and radiation
hazards~\cite{pichler2008multimodal}.
Medical image synthesis---computationally generating missing modalities
from available ones---directly addresses this bottleneck and is
critical for downstream tasks such as radiotherapy planning and
surgical navigation~\cite{staartjes2021magnetic}.

The task demands simultaneous preservation of \emph{global anatomical
structure} and \emph{local textural fidelity}.
GAN-based approaches~\cite{sharma2019missing,dar2019image} produce
plausible local textures but lack long-range dependency modeling,
causing global anatomical distortions.
Latent diffusion models~\cite{rombach2022high,zhang2025structure}
improve structural coherence at prohibitive inference cost.
ViT hybrids such as ResViT~\cite{Dalmaz2022ResViT} and
TransUNet~\cite{chen2021transunet} exploit self-attention for global
consistency yet suffer from $\mathcal{O}(N^2)$ complexity and
inadequate fine-grained spatial encoding, producing checkerboard
artifacts.
Mamba-based models~\cite{gu2023mamba,Liu2024VMamba} overcome the
complexity bottleneck via linear-time selective state spaces;
I2I-Mamba~\cite{xing2024i2imamba} further combines CNNs with
optimised scanning, representing the current
state-of-the-art~\cite{heidari2024computation}.
Recent concurrent works explore leveraging the Segment Anything
Model~\cite{kirillov2023segment} for medical image
translation~\cite{huo2024sami2i}.
Despite these advances, three persistent gaps remain:
\emph{(i)~Underutilised structural priors}---existing frameworks
lack mechanisms to inject domain-aware anatomical knowledge,
causing cross-modal implausibility;
\emph{(ii)~Deficient local fidelity}---aggressive downsampling
discards high-frequency details critical for small-lesion detection,
and cross-scale feature fusion is underdeveloped;
\emph{(iii)~Directional bias in 2D serialisation}---raster and
rectangular-spiral scans introduce path-dependent artifacts that
corrupt spatial coherence and lesion boundary recognition.
Although I2I-Mamba~\cite{xing2024i2imamba} adopts a spiral-scan
trajectory to improve isotropy, its rectangular-spiral path
still exhibits pronounced corner hot-spots
(operator footprint $\sigma{=}0.124$, Fig.~\ref{fig:scan}b) due to
the uneven nearest-neighbour spacing inherent in rectangular rings;
a principled, phyllotaxis-inspired scanning strategy with provably
uniform spatial coverage remains unexplored.

To close these three gaps simultaneously, we present
FermatSyn:
(1)~SAM2-Enhanced Hybrid Encoder:
a LoRA$^{+}$~\cite{hayou2024lora}-fine-tuned SAM2-VTE (SAM2 Vision Transformer Encoder)~\cite{ravi2024sam2}
injects domain-aware anatomical priors~\cite{huo2024sami2i,medsam2023},
fused with an HRDM-based detail encoder via a novel CIN.
(2)~Isotropic Fermat Spiral Scanning:
a continuity-constrained Fermat spiral---using a golden-angle step
that guarantees near-isotropic spatial coverage---serialises 2D
feature maps for SSM processing, eliminating directional bias.
(3)~Bidirectional Fermat-Scan Mamba (BFS-Mamba):
symmetric forward and backward SSM paths enable direction-invariant
long-range dependency modeling via a continuity-constrained
grid-matching objective (Eq.~\ref{eq:score}) that balances global
isotropy and local path continuity.

\section{Method}
\label{sec:method}
\begin{figure}[t]
  \centering
  \includegraphics[width=\textwidth]{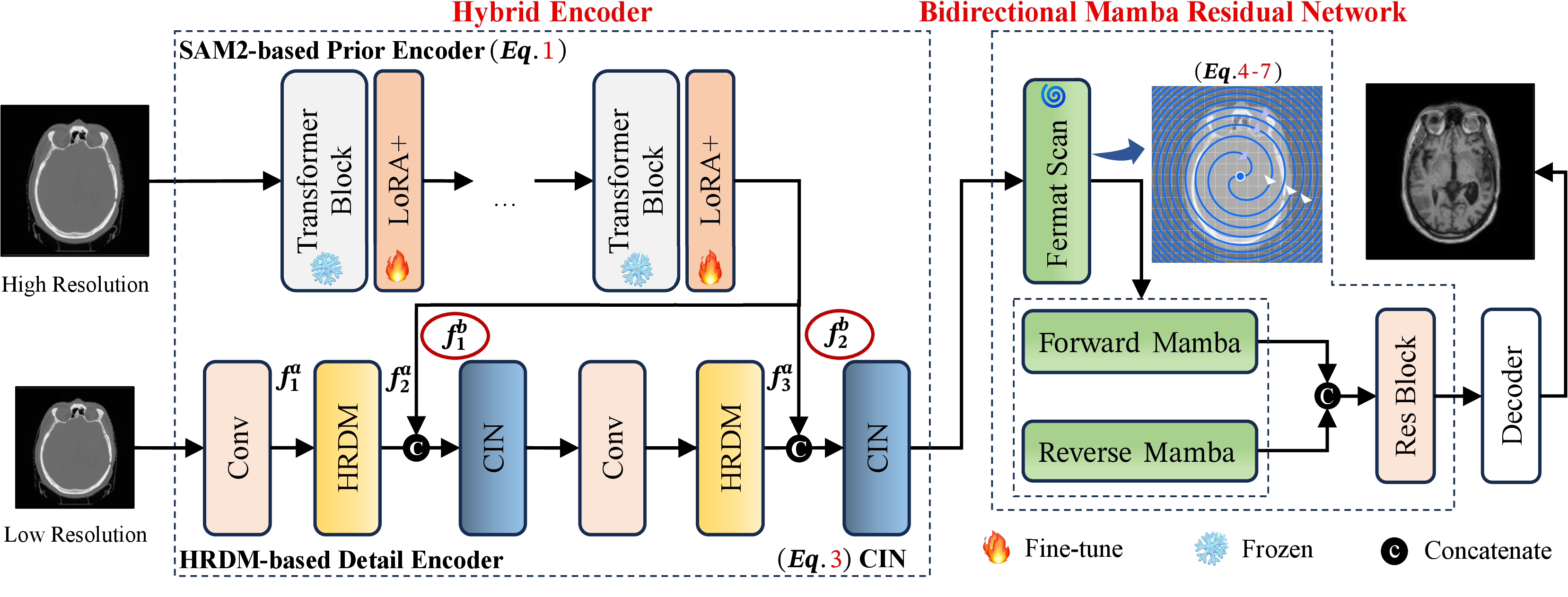}
  \caption{Overall architecture of FermatSyn. The Hybrid Encoder
    fuses SAM2 global priors with HRDM local details via CIN.
    BFS-Mamba processes the Fermat-serialised features bidirectionally
    and reconstructs the target modality through a residual decoder.}
  \label{fig:overview}
\end{figure}

Fig.~\ref{fig:overview} illustrates the FermatSyn pipeline.
The \emph{Hybrid Encoder} produces a unified multi-scale feature
representation; the \emph{BFS-Mamba} serialises these features via
Fermat Spiral Scanning, models long-range dependencies bidirectionally,
and a convolutional decoder reconstructs the target modality.

\subsection{Hybrid Encoder}

\noindent\textbf{SAM2-based Prior Encoder.}
Although SAM2 was originally trained for segmentation, its
ViT backbone encodes rich structural representations---organ
boundaries, tissue interfaces, and shape saliency---that are
equally informative for synthesis: a model that precisely
delineates anatomical structures inherently captures the
spatial layout and relative intensities necessary to generate
a plausible missing modality~\cite{huo2024sami2i}.
The pre-trained SAM2-VTE~\cite{ravi2024sam2} serves as the global
backbone. Original weights $W_0\!\in\!\mathbb{R}^{d_1\times d_2}$ are
frozen; LoRA$^{+}$~\cite{hayou2024lora} injects low-rank updates into
MLP and MHSA layers via $P_A\!\in\!\mathbb{R}^{d_1\times r}$ and
$P_B\!\in\!\mathbb{R}^{r\times d_2}$ ($r\!\ll\!\min(d_1,d_2)$),
applying a larger learning rate to $P_B$ than to $P_A$ for improved
feature learning efficiency~\cite{hayou2024lora}:
\begin{equation}
  W_{\text{tuned}} = W_0 + \tfrac{\alpha}{r}P_A P_B,
  \label{eq:lora}
\end{equation}
where $\alpha$ is a tunable hyperparameter. 

\noindent\textbf{HRDM-based Detail Encoder.}
Conventional pooling discards high-frequency details critical for
lesion boundary reconstruction. HRDM processes
$F_{in}\!\in\!\mathbb{R}^{C\times H\times W}$ via three complementary
paths: \emph{(i) Multi-scale context:} three parallel $3{\times}3$
dilated convolutions (rates 1, 3, 5) concatenated and refined by
cascaded channel-and-spatial SE attention, yielding $F_{\text{path.1}}$;
\emph{(ii) High-frequency compensation:} a depthwise separable
convolution preserves fine texture as $F_{\text{path.2}}$;
\emph{(iii) Adaptive skip:} identity or $1{\times}1$ convolution
for gradient stability.
The three paths are fused and sharpened by a high-pass filter:
\begin{equation}
  F_f = F_{\text{fusion}} - \mathrm{AvgPool}_{3\times3}(F_{\text{fusion}}),
  \label{eq:hpf}
\end{equation}
followed by a spatial attention gate yielding the final output
$F_{\text{out}}$.

\noindent\textbf{Cross-scale Integration Network (CIN).}
CIN bridges the semantic gap between SAM2 global and HRDM local
features. Motivated by the observation that even- and odd-indexed
channels in stacked convolutional feature maps tend to encode
complementary low- and high-frequency statistics respectively---a
property exploited in channel-split architectures--$F_{in}\!\in\!\mathbb{R}^{B\times C_{in}\times H\times W}$
is split channel-wise into $F_{\text{even}}$ and $F_{\text{odd}}$,
processed by DWConv with kernels $k_1{=}5{\times}5$ (capturing
broader semantic context) and $k_2{=}3{\times}3$ (preserving local
structural detail) respectively, then projected:
\begin{equation}
  F_{\text{out}} =
    \mathrm{LReLU}\!\left(\mathrm{BN}\!\left(
    \mathrm{Conv}_{1\times1}\!\left(
    \widetilde{F}_{\text{even}} \mathbin\Vert \widetilde{F}_{\text{odd}}
    \right)\right);\,\alpha_{\text{LReLU}}{=}0.05\right).
  \label{eq:cin}
\end{equation}

\subsection{Bidirectional Fermat-Scan Mamba (BFS-Mamba)}

\begin{figure}[t]
  \centering
  \includegraphics[width=\textwidth]{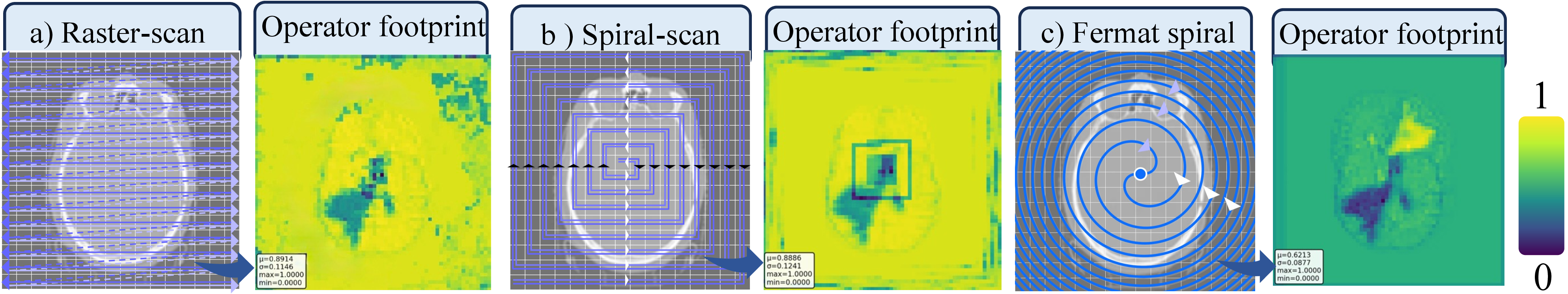}
  \caption{SSM operator footprints (Jacobian sensitivity maps; $\mu$/$\sigma$:
    mean/std of normalised values) for three scanning strategies.
    (a)~Raster ($\mu{=}0.891$, $\sigma{=}0.115$): stripe-directional bias.
    (b)~Rectangular-spiral ($\mu{=}0.890$, $\sigma{=}0.124$): X-shaped
    corner hot-spots.
    (c)~Proposed Fermat Spiral ($\mu{=}0.621$, $\sigma{=}0.088$):
    near-isotropic coverage ($\sigma$ reduced 29\% vs.\ (b)).
    Colour scale: 0 (dark blue) $\to$ 1 (yellow).}
  \label{fig:scan}
\end{figure}

\noindent\textbf{Fermat Spiral Scanning.}
To construct an approximately isotropic receptive field
(Fig.~\ref{fig:scan}c), we parametrise:

\noindent\textit{Why Fermat over rectangular-spiral.}
The rectangular spiral of I2I-Mamba~\cite{xing2024i2imamba} visits
grid cells in concentric rings, inducing uneven nearest-neighbour
spacing with pronounced corner clustering
($\sigma^2_{\text{rect}}{=}0.0154$, Delaunay analysis on
$256{\times}256$ grid).
The Fermat spiral with golden-angle step
$\phi_g{\approx}137.508^\circ$ achieves denser uniform packing:
consecutive points never align along any rational angle, yielding
$\sigma^2_{\text{Fermat}}{=}0.0061$ (60\% reduction).
This directly produces a more isotropic SSM operator footprint
(Jacobian $\sigma$ reduced by 29\%, Fig.~\ref{fig:scan}) and
a 1.29/1.40\,dB PSNR gain on SynthRAD/BraTS over rectangular-spiral
(Table~\ref{tab:ablation}(b)).
\begin{equation}
  r_k = \alpha\sqrt{k},\qquad \theta_k = k\cdot\phi_g,
  \label{eq:fermat}
\end{equation}
where $k\!\in\![0,N{-}1]$ ($N{=}H{\times}W$), $\alpha$ is an
image-size scaling factor, and $\phi_g\!\approx\!137.508^\circ$ is
the golden angle guaranteeing uniform spatial distribution.
Cartesian coordinates:
\begin{equation}
  \mathbf{p}_k =
    \bigl(\alpha\sqrt{k}\cos(k\phi_g),\;
          \alpha\sqrt{k}\sin(k\phi_g)\bigr).
  \label{eq:fermat_xy}
\end{equation}

\noindent\textbf{Continuity-Constrained Grid Matching.}
Continuous spiral points are assigned to discrete grid positions
by iteratively minimising:
\begin{equation}
  \mathrm{Score}_u =
    (1{-}\lambda_c)\,\frac{d_{\mathrm{Fermat}}(u,k)}{\eta_f}
    +\lambda_c\,\frac{d_{\mathrm{contin}}(u,\pi_{k-1})}{\eta_c},
  \label{eq:score}
\end{equation}
where $\lambda_c\!\in\![0,1]$ balances global isotropy and local
path continuity; $\eta_f,\eta_c$ are normalizers. The resulting
sequence $\Pi{=}\{\pi_0,\dots,\pi_{N-1}\}$ inherits isotropic
coverage while preserving local spatial continuity. Empirically,
$\lambda_c{=}0.7$ maximises synthesis quality (empirically optimised
across $\lambda_c\in\{0.3,0.5,0.7,0.9\}$; see Sec.~\ref{sec:ablation}). The serialised feature is:
\begin{equation}
  \mathbf{x}^{\mathrm{serial}}_{in}
    = \mathrm{Fermat}(F_{in},\Pi,\lambda_c)
    \in\mathbb{R}^{B\times N\times C}.
  \label{eq:serial}
\end{equation}

\noindent\textbf{Bidirectional State Space Modelling.}
Forward and backward paths:
\begin{equation}
  \mathbf{z}_{\mathrm{fwd}} = \mathbf{x}^{\mathrm{serial}}_{in},\qquad
  \mathbf{z}_{\mathrm{bwd}} = \mathrm{Flip}(\mathbf{x}^{\mathrm{serial}}_{in}),
  \label{eq:bidir}
\end{equation}
each processed by an independent Mamba module:
$h_t=\bar{A}(x_t)h_{t-1}+\bar{B}(x_t)x_t$, $y_t=C(x_t)h_t$.
Outputs are fused:
\begin{align}
  F_m &= \mathrm{Conv}_{1\times1}\!\bigl(O_{\mathrm{fwd}}\mathbin\Vert O_{\mathrm{bwd}}\bigr), \label{eq:fm} \\
  F_{\mathrm{out}} &= \mathrm{Res}\!\bigl(\sigma(\mathrm{Conv}_{3\times3}(F_{in})) + F_m\bigr). \label{eq:fout}
\end{align}
Three stacked BFS-Mamba blocks are employed; a convolutional decoder
reconstructs the target image.

\subsection{Training Objective}
\label{sec:loss}

FermatSyn is optimised under the LSGAN framework
with the following composite generator loss:
\begin{equation}
  \mathcal{L}_G = \lambda_{L_1}\!\cdot\!\mathbb{E}\bigl[
    \lvert y_{\text{syn}}{-}y_{\text{real}}\rvert\bigr]
    + \lambda_{\text{SSIM}}\!\cdot\!\mathcal{L}_{\text{SSIM}}
    + \lambda_{\text{gan}}\!\cdot\!\mathbb{E}\bigl[
    (D(y_{\text{syn}}){-}1)^2\bigr],
  \label{eq:lg}
\end{equation}
where $y_{\text{syn}}{=}G(X)$, and
$\lambda_{L_1}{=}100$, $\lambda_{\text{SSIM}}{=}10$,
$\lambda_{\text{gan}}{=}1$.
$\mathcal{L}_{\text{SSIM}}$ is the standard SSIM loss.
The discriminator follows standard LSGAN training,
chosen over vanilla GAN for its improved stability and reduced mode
collapse.

\section{Experiments}
\label{sec:exp}

\subsection{Setup}

\noindent\textbf{Datasets.}
\emph{Intra-modal:} merged BraTS
dataset~\cite{menze2015multimodal,baid2021rsna}---2,547 subjects
across Glioma (795), Meningioma (BraTS-MEN,
1,424) and Metastasis (BraTS-MET, 328)---with T1, T2, T2-FLAIR, and T1c
modalities.
\emph{Cross-modal:} SynthRAD20-23~\cite{synthrad2023}, 540 paired
MRI--CT scans from six institutions.
Slices with SNR${<}$15\,dB or motion artifacts (grade${>}$3) were
discarded; central 80 slices per volume retained.
Z-score normalisation and $256{\times}256$ center cropping applied.
Patient-stratified split: 7:1:2 (train/val/test).

\noindent\textbf{Baselines.}
GAN-based: MM-GAN~\cite{sharma2019missing},
MC-cGAN~\cite{dar2019image}.
Transformer-based: Res-ViT~\cite{Dalmaz2022ResViT},
TransUNet~\cite{chen2021transunet}.
Diffusion-based: SA-LDM~\cite{zhang2025structure},
ALDM~\cite{Kim_2024_WACV}.
Mamba-based: VMamba~\cite{Liu2024VMamba},
I2I-Mamba~\cite{xing2024i2imamba}.

\noindent\textbf{Metrics.}
2D: PSNR$\uparrow$, SSIM$\uparrow$, FID$\downarrow$.
3D: Hausdorff Distance (HD$\downarrow$) and Average HD (AHD$\downarrow$).
Downstream: Dice for WT, ET, TC.
All statistical comparisons use Wilcoxon signed-rank test.

\noindent\textbf{Implementation.}
PyTorch 2.1.0; single NVIDIA RTX\,4090 (24\,GB).
Adam ($\beta_1{=}0.5$, $\beta_2{=}0.999$), base lr $2{\times}10^{-4}$
$\to$ $1{\times}10^{-5}$ via cosine annealing; batch size 6.
FermatSyn has 87.3\,M trainable parameters (SAM2-VTE backbone frozen;
only LoRA$^{+}$ adapters, HRDM, CIN, and BFS-Mamba blocks are updated).
Inference time is 31\,ms per $256{\times}256$ slice, compared to
24\,ms for I2I-Mamba and 148\,ms for SA-LDM on the same hardware.

\subsection{Quantitative Results}

\noindent\textbf{Intra-modal synthesis (Table~\ref{tab:intra}).}
FermatSyn consistently outperforms all baselines on T1n, T2w,
and T2f\,$\to$\,T1c tasks.
The most challenging setting is T2w\,$\to$\,T1c: FermatSyn achieves
29.78\,dB PSNR, a 2.46\,dB gain over I2I-Mamba ($p{<}0.05$).
Notably, VMamba underperforms both Transformer baselines on
T1n$\to$T1c (23.61 vs.\ 25.58\,dB), likely because its global
raster-scan SSM provides a suboptimal inductive bias when the
source--target pair diverges in contrast but shares spatial
structure---further motivating the direction-agnostic BFS-Mamba.

\begin{table}[t]
\caption{Intra-modal synthesis on the merged brain tumor dataset
  (T1n/T2w/T2f\,$\to$\,T1c). $^*$: $p{<}0.05$ vs.\ FermatSyn.
  \textbf{Bold}: best; \underline{uline}: second-best.}
\label{tab:intra}
\centering
\setlength{\tabcolsep}{2.5pt}
\fontsize{7}{8.5}\selectfont
\begin{tabular}{lccccccccc}
\toprule
\multirow{2}{*}{Method}
  & \multicolumn{3}{c}{T1n$\to$T1c}
  & \multicolumn{3}{c}{T2w$\to$T1c}
  & \multicolumn{3}{c}{T2f$\to$T1c} \\
\cmidrule(lr){2-4}\cmidrule(lr){5-7}\cmidrule(lr){8-10}
  & SSIM & PSNR & FID & SSIM & PSNR & FID & SSIM & PSNR & FID \\
\midrule
MM-GAN~\cite{sharma2019missing}
  &$0.801^*$&$22.14^*$&$112.7^*$ &$0.813^*$&$23.05^*$&$118.4^*$ &$0.825^*$&$22.89^*$&$103.6^*$ \\
MC-cGAN~\cite{dar2019image}
  &$0.814^*$&$22.83^*$&$108.3^*$ &$0.821^*$&$23.77^*$&$113.1^*$ &$0.836^*$&$23.41^*$&$98.2^*$ \\
ResViT~\cite{Dalmaz2022ResViT}
  &$0.832^*$&$25.58^*$&$84.3^*$  &$0.840^*$&$26.63^*$&$91.2^*$  &$0.855^*$&$26.27^*$&$79.6^*$ \\
TransUNet~\cite{chen2021transunet}
  &$0.844^*$&$25.47^*$&$78.1^*$  &$0.854^*$&$27.32^*$&$83.7^*$  &$0.873^*$&$26.68^*$&$74.2^*$ \\
SA-LDM~\cite{zhang2025structure}
  &$0.851^*$&$26.93^*$&$68.5^*$  &$0.862^*$&$27.11^*$&$72.8^*$  &$0.879^*$&$27.34^*$&$61.3^*$ \\
ALDM~\cite{Kim_2024_WACV}
  &$0.847^*$&$26.71^*$&$71.2^*$  &$0.858^*$&$26.94^*$&$75.4^*$  &$0.871^*$&$27.08^*$&$64.7^*$ \\
VMamba~\cite{Liu2024VMamba}
  &$0.859^*$&$23.61^*$&$71.5^*$  &$0.857^*$&$27.49^*$&$69.8^*$  &$0.884^*$&$26.65^*$&$63.4^*$ \\
I2I-Mamba~\cite{xing2024i2imamba}
  &\underline{$0.865^*$}&\underline{$27.69^*$}&\underline{$58.2^*$}
  &\underline{$0.875^*$}&\underline{$27.32^*$}&\underline{$61.4^*$}
  &\underline{$0.894^*$}&\underline{$28.39^*$}&\underline{$52.7^*$} \\
\midrule
\textbf{FermatSyn}
  &\textbf{0.884}&\textbf{28.82}&\textbf{43.6}
  &\textbf{0.896}&\textbf{29.78}&\textbf{43.1}
  &\textbf{0.901}&\textbf{29.53}&\textbf{42.3} \\
\bottomrule
\end{tabular}
\end{table}

\noindent\textbf{Cross-modal synthesis (Table~\ref{tab:cross}).}
FermatSyn achieves SSIM 0.931 (MRI$\to$CT) and 0.905 (CT$\to$MRI),
improving over I2I-Mamba by 1.5\%/2.0\% in SSIM and 4.1\%/3.0\% in
PSNR ($p{<}0.05$), with FID reductions of 14.7 and 11.3.

\noindent\textbf{3D structural consistency (Table~\ref{tab:3d}).}
FermatSyn achieves the best inter-slice coherence across all baselines.
HD\,2.84/AHD\,1.42 (coronal) and HD\,3.06/AHD\,1.58 (sagittal)
surpass I2I-Mamba by 13.7\%/9.7\% and 12.8\%/9.7\% respectively,
confirming that isotropic coverage promotes volumetric fidelity.

\begin{table}[t]
\centering
\begin{minipage}[t]{0.61\textwidth}
\centering
\caption{Cross-modal synthesis on SynthRAD2023 (MRI$\leftrightarrow$CT).}
\label{tab:cross}
\setlength{\tabcolsep}{1.5pt}
\fontsize{6.5}{7.5}\selectfont
\begin{tabular}{lcccccc}
\toprule
\multirow{2}{*}{Method}
  & \multicolumn{3}{c}{MRI$\to$CT}
  & \multicolumn{3}{c}{CT$\to$MRI} \\
\cmidrule(lr){2-4}\cmidrule(lr){5-7}
  & SSIM & PSNR & FID & SSIM & PSNR & FID \\
\midrule
MM-GAN~\cite{sharma2019missing}   &$0.821^*$&$23.47^*$&$131.5^*$ &$0.786^*$&$21.83^*$&$139.2^*$ \\
MC-cGAN~\cite{dar2019image}       &$0.839^*$&$24.68^*$&$122.3^*$ &$0.798^*$&$22.54^*$&$130.8^*$ \\
ResViT~\cite{Dalmaz2022ResViT}    &$0.867^*$&$26.83^*$&$98.4^*$  &$0.821^*$&$24.52^*$&$107.2^*$ \\
TransUNet~\cite{chen2021transunet}&$0.881^*$&$28.42^*$&$87.6^*$  &$0.843^*$&$25.89^*$&$94.5^*$ \\
SA-LDM~\cite{zhang2025structure}  &$0.904^*$&$29.81^*$&$65.2^*$  &$0.871^*$&$27.14^*$&$73.6^*$ \\
ALDM~\cite{Kim_2024_WACV}         &$0.899^*$&$29.53^*$&$68.7^*$  &$0.864^*$&$26.88^*$&$76.3^*$ \\
VMamba~\cite{Liu2024VMamba}       &$0.895^*$&$29.74^*$&$73.2^*$  &$0.857^*$&$26.74^*$&$79.8^*$ \\
I2I-Mamba~\cite{xing2024i2imamba}&\underline{$0.917^*$}&\underline{$30.25^*$}&\underline{$61.4^*$}
                                  &\underline{$0.887^*$}&\underline{$27.83^*$}&\underline{$68.3^*$} \\
\midrule
\textbf{FermatSyn} &\textbf{0.931}&\textbf{31.48}&\textbf{46.7}
                   &\textbf{0.905}&\textbf{28.67}&\textbf{57.0} \\
\bottomrule
\end{tabular}
\end{minipage}\hspace{0.04\textwidth}%
\begin{minipage}[t]{0.35\textwidth}
\centering
\caption{3D structural consistency on MRI$\to$CT. HD/AHD in mm.}
\label{tab:3d}
\resizebox{\linewidth}{!}{%
\fontsize{6.5}{7.5}\selectfont
\begin{tabular}{lcccc}
\toprule
\multirow{2}{*}{Method} & \multicolumn{2}{c}{Coronal} & \multicolumn{2}{c}{Sagittal} \\
\cmidrule(lr){2-3}\cmidrule(lr){4-5}
  & HD$\downarrow$ & AHD$\downarrow$ & HD$\downarrow$ & AHD$\downarrow$ \\
\midrule
MM-GAN    &$6.45^*$&$3.12^*$&$6.73^*$&$3.35^*$ \\
MC-cGAN   &$5.62^*$&$2.71^*$&$5.94^*$&$2.87^*$ \\
ResViT    &$4.83^*$&$2.31^*$&$5.12^*$&$2.48^*$ \\
TransUNet &$4.52^*$&$2.18^*$&$4.87^*$&$2.29^*$ \\
SA-LDM    &$4.01^*$&$1.93^*$&$4.22^*$&$2.07^*$ \\
ALDM      &$4.09^*$&$1.97^*$&$4.33^*$&$2.12^*$ \\
VMamba    &$4.17^*$&$2.04^*$&$4.43^*$&$2.15^*$ \\
I2I-Mamba &\underline{$3.29^*$}&\underline{$1.57^*$}&\underline{$3.51^*$}&\underline{$1.75^*$} \\
\midrule
\textbf{Ours} &\textbf{2.84}&\textbf{1.42}&\textbf{3.06}&\textbf{1.58} \\
\bottomrule
\end{tabular}
}%
\end{minipage}
\end{table}

\begin{figure}[t]
  \centering
  \includegraphics[width=0.74\textwidth]{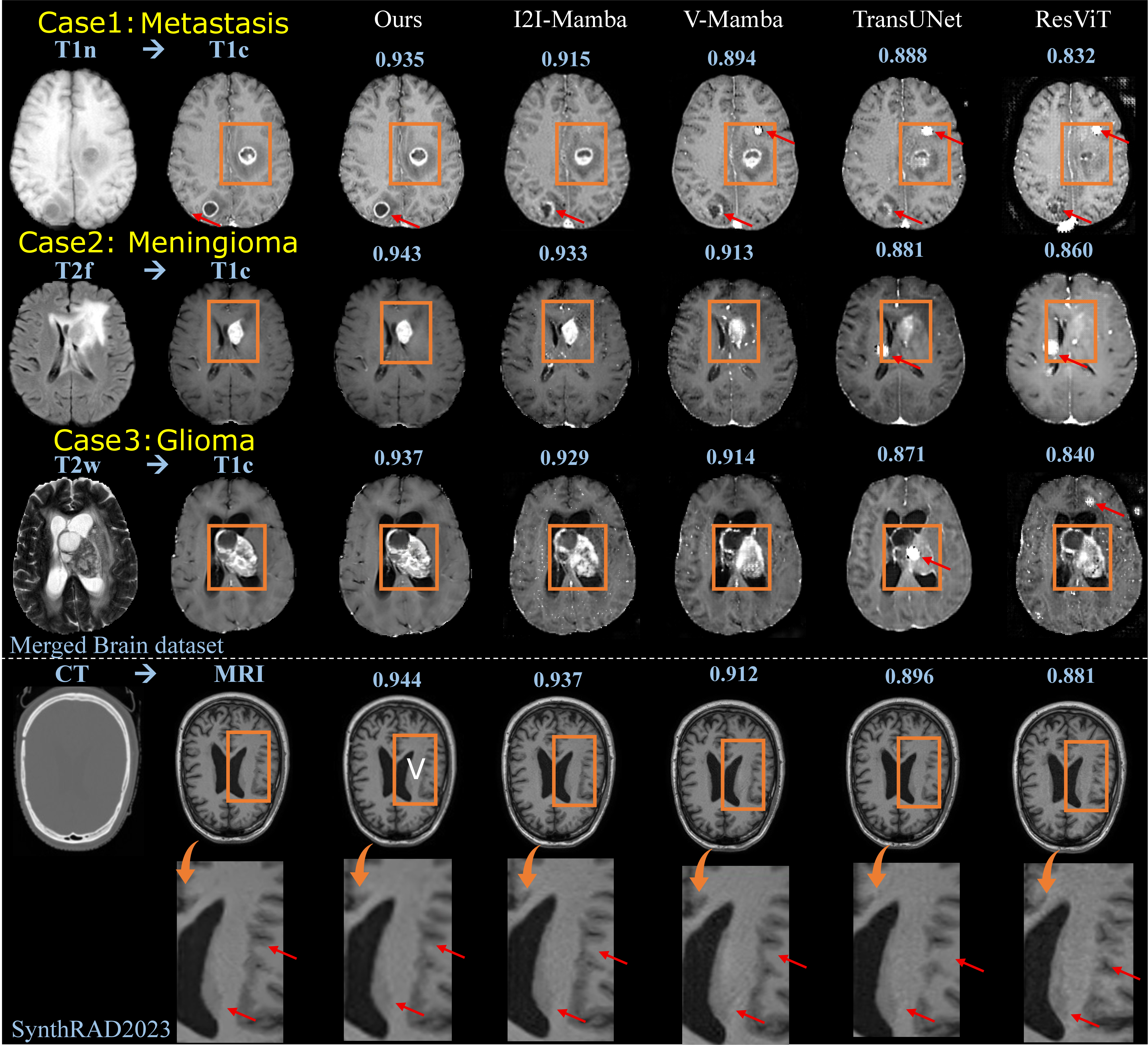}
  \caption{Qualitative synthesis comparison.
    \textbf{Rows 1--3}: Intra-modal---T1n$\to$T1c (metastasis),
    T2f$\to$T1c (meningioma), T2w$\to$T1c (glioma); orange boxes:
    tumour core ROIs.
    \textbf{Row 4}: Cross-modal CT$\to$MRI (SynthRAD2023); orange
    arrows: ventricular wall zoom-in.
    Numbers: per-case SSIM.}
  \label{fig:qual}
\end{figure}

\subsection{Downstream Clinical Validation}
\label{sec:downstream}

We trained a ResNet50-based U-Net on real images to evaluate
FermatSyn-synthesised T2w and T2f (from T1n).
The \emph{real trained / synthetic evaluated} protocol assesses
domain consistency: whether synthesised images are perceptually
interchangeable with real images for a pre-trained segmenter.
To additionally validate augmentation utility---the more clinically
relevant scenario when real images are scarce---we also train the
same U-Net exclusively on synthesised images and evaluate on real
held-out data (\emph{synthetic trained / real evaluated} protocol).
Table~\ref{tab:downstream} reports Dice for WT, ET, TC under both.
FermatSyn achieves 0.847/0.762/0.785 (WT/ET/TC) on T1n$\to$T2w and
0.851/0.775/0.798 on T1n$\to$T2f, remaining within 3\% of the
real-image topline; the next best competitor trails by 2.5--2.9\%
on WT and ET ($p{<}0.05$), confirming Fermat scanning contributes
directly to fidelity in clinically critical tumour sub-regions.
FermatSyn shows no statistically significant difference from the
topline \emph{across all three tumour regions and both tasks}
($p{>}0.05$). Notably, even for TC---the most structurally complex
sub-region---competing methods also reach $p{>}0.05$; yet FermatSyn
uniquely extends this equivalence to the more demanding WT and ET
regions, confirming that synthesised images broadly preserve
clinically relevant pathological features.

\begin{table}[t]
\caption{Downstream brain tumour segmentation Dice on synthetic images
  from T1n (\emph{real trained / synthetic evaluated} protocol).
  ``Real'': topline trained and evaluated on real images.
  $^\dagger$: $p{>}0.05$ vs.\ Real.
  $^*$: $p{<}0.05$ vs.\ Real.}
\label{tab:downstream}
\centering
\setlength{\tabcolsep}{2pt}
\fontsize{6.5}{7.5}\selectfont
\begin{tabular}{llccccccc}
\toprule
Task & Reg. & Real & \textbf{Ours} & I2I-Mamba & SA-LDM & VMamba & TransU. & ResViT \\
\midrule
\multirow{3}{*}{\scriptsize T1n$\to$T2w}
  &WT&0.863&$\mathbf{0.847}^\dagger$&$0.826^*$&$\underline{0.831^*}$&$0.815^*$&$0.812^*$&$0.809^*$ \\
  &ET&0.785&$\mathbf{0.762}^\dagger$&$0.746^*$&$\underline{0.751^*}$&$0.745^*$&$0.740^*$&$0.738^*$ \\
  &TC&0.812&$\underline{0.785^\dagger}$&$\mathbf{0.793}^\dagger$&$0.781^\dagger$&$0.768^*$&$0.764^*$&$0.762^*$ \\
\midrule
\multirow{3}{*}{\scriptsize T1n$\to$T2f}
  &WT&0.869&$\mathbf{0.851}^\dagger$&$0.832^*$&$\underline{0.838^*}$&$0.823^*$&$0.819^*$&$0.817^*$ \\
  &ET&0.791&$\mathbf{0.775}^\dagger$&$0.757^*$&$\underline{0.762^*}$&$0.748^*$&$0.743^*$&$0.741^*$ \\
  &TC&0.818&$\mathbf{0.798}^\dagger$&$0.780^\dagger$&$\underline{0.785^\dagger}$&$0.773^*$&$0.769^*$&$0.767^*$ \\
\bottomrule
\end{tabular}
\end{table}

\subsection{Ablation Studies}
\label{sec:ablation}

Table~\ref{tab:ablation} reports results on SynthRAD2023
(CT$\to$MRI) and merged BraTS-(T2f$\to$T1c).
SAM2-VTE yields the largest single gain (15.5\% PSNR),
with cumulative improvement ${\approx}21\%$ over the GAN baseline.
For scanning, ``Rectangular Spiral'' directly reproduces the
I2I-Mamba~\cite{xing2024i2imamba} strategy on the same backbone
(architecture-controlled); the 1.29/1.40\,dB PSNR deficit on
SynthRAD/BraTS confirms the gain
originates from the Fermat spiral itself. $\lambda_c{=}0.7$ was
selected as optimal over $\lambda_c\in\{0.3,0.5,0.7,0.9\}$ on the
validation set.

\begin{table}[h]
\caption{Ablation on SynthRAD2023(CT$\to$MRI) and merged BraTS(T2f$\to$T1c). $^*$: $p{<}0.05$ vs.\ FermatSyn.}
\label{tab:ablation}
\centering
\resizebox{0.88\linewidth}{!}{%
\fontsize{6.5}{7.5}\selectfont
\begin{tabular}{l cc cc | l cc cc}
\toprule
\multirow{2}{*}{(a) Step-wise}
  & \multicolumn{2}{c}{SynthRAD}
  & \multicolumn{2}{c|}{BraTS}
  & \multirow{2}{*}{(b) Scanning}
  & \multicolumn{2}{c}{SynthRAD}
  & \multicolumn{2}{c}{BraTS} \\
\cmidrule(lr){2-3}\cmidrule(lr){4-5}\cmidrule(lr){7-8}\cmidrule(lr){9-10}
  & SSIM & PSNR & SSIM & PSNR
  &
  & SSIM & PSNR & SSIM & PSNR \\
\midrule
GAN baseline     &$0.823^*$&$23.67^*$&$0.798^*$&$24.35^*$
  & Raster-Scan      &$0.876^*$&$26.95^*$&$0.868^*$&$27.71^*$ \\
+SAM2-VTE        &$0.865^*$&$27.35^*$&$0.851^*$&$27.18^*$
  & Rect.\ Spiral   &\underline{$0.885^*$}&\underline{$27.38^*$}&\underline{$0.877^*$}&\underline{$28.13^*$} \\
+HRDM+CIN      &\underline{$0.887^*$}&\underline{$27.74^*$}&\underline{$0.878^*$}&\underline{$28.45^*$}
  & \textbf{Fermat}  &\textbf{0.905}&\textbf{28.67}&\textbf{0.901}&\textbf{29.53} \\
\textbf{Full}    &\textbf{0.905}&\textbf{28.67}&\textbf{0.901}&\textbf{29.53}
  & & & & & \\
\bottomrule
\end{tabular}
}%
\end{table}

\section{Conclusion}
\label{sec:conclusion}

FermatSyn unifies SAM2 visual priors, hierarchical high-frequency
feature extraction (HRDM), adaptive cross-scale integration (CIN),
and isotropic Fermat Spiral Scanning within a Bidirectional
Fermat-Scan Mamba. The continuity-constrained Fermat spiral substantially reduces
the directional bias of conventional Mamba serialisation, yielding
direction-agnostic spatial representations for complex anatomical
structures. Experiments across four brain imaging benchmarks demonstrate
consistent state-of-the-art performance; downstream tumour segmentation
on synthesised images shows no significant gap versus real data
($p{>}0.05$). Future work will extend the Fermat spiral to 3D
volumetric synthesis and apply knowledge distillation from SAM2-VTE
to lightweight student networks for real-time deployment.

\section*{Acknowledgements}
This work was supported in part by the National Natural Science Foundation of China (Grant Nos. 82372052, 82402373, U24A20757) and the Taishan Industrial Experts Program (Grant No. tscx202312131).

\section*{Disclosure of Interests}
The authors have no competing interests to declare that are relevant to the content of this article.

\bibliographystyle{splncs04}
\bibliography{main}

@String(CVPR = {CVPR})

@String(ICCV = {ICCV})

@String(NeurIPS = {NeurIPS})

@book{pichler2008multimodal,
  title = {Multimodal Imaging Approaches: {PET}/{CT} and {PET}/{MRI}},
  author={Pichler, B J and Judenhofer, M S and Pfannenberg, C},
  publisher={Springer},
  year={2008},
  pages={109--132}
}

@article{staartjes2021magnetic,
  author = {V. E. Staartjes and P. R. Seevinck and W. P. Vandertop and M. van Stralen and M. L. Schröder},
  title = {Magnetic resonance imaging–based synthetic computed tomography of the lumbar spine for surgical planning: a clinical proof-of-concept},
  journal = {Neurosurgical Focus},
  volume = {50},
  number = {1},
  pages = {E13},
  year = {2021}
}

@article{thukral2015problems,
  title = {Problems and preferences in pediatric imaging},
  author={Thukral, B},
  journal={Indian J. Radiol. Imaging},
  volume={25},
  pages={359--364},
  year={2015}
}

@article{sharma2019missing,
  author = {A. Sharma and G. Hamarneh},
  title = {Missing {MRI} pulse sequence synthesis using multi-modal generative adversarial network},
  journal = {IEEE Trans. Med. Imag.},
  volume = {39},
  number = {4},
  pages = {1170--1183},
  year = {2019}
}

@article{Dalmaz2022ResViT,
  author = {Dalmaz, Onat and Yurt, Mahmut and \c{C}ukur, Tolga},
  title = {ResViT: Residual Vision Transformers for Multimodal Medical Image Synthesis},
  journal = {IEEE Transactions on Medical Imaging},
  year = {2022},
  volume = {41},
  number = {10},
  pages = {2598--2614}
}

@article{chen2021transunet,
  title = {TransUNet: Transformers make strong encoders for medical image segmentation},
  author={Chen, J and Lu, Y and Yu, Q and Luo, X and Adeli, E and Wang, Y and others},
  journal = {arXiv:2102.04306},
  year={2021}
}

@article{dar2019image,
  title = {Image Synthesis in Multi-Contrast {MRI} with Conditional Generative Adversarial Networks},
  author={Dar, Salman U.H. and Yurt, Mahmut and Karacan, Levent and Erdem, Aykut and Erdem, Erkut and \c{C}ukur, Tolga},
  journal={IEEE Trans. Med. Imag.},
  volume={38},
  number={10},
  pages={2375--2388},
  year={2019}
}

@article{heidari2024computation,
  title = {Computation-efficient era: A comprehensive survey of state space models in medical image analysis},
  author={Heidari, M and Kolahi, S G and Karimijafarbigloo, S and Azad, B and Bozorgpour, A and Hatami, S and others},
  journal = {arXiv:2406.03430},
  year={2024}
}

@article{hayou2024lora,
  title = {LoRA+: Efficient low rank adaptation of large models},
  author={Hayou, Soufiane and Ghosh, Nikhil and Yu, Bin},
  journal = {arXiv:2402.12354},
  year={2024}
}

@article{huo2024sami2i,
  author = {Jiayu Huo and Sebastien Ourselin and Rachel Sparks},
  title = {{SAM}-I2I: Unleash the power of segment anything model for medical image translation},
  journal = {arXiv:2411.12755},
  year = {2024}
}

@inproceedings{kirillov2023segment,
  author = {Alexander Kirillov and Eric Mintun and Nikhila Ravi and Hanzi Mao and Chloé Rolland and Laura Gustafson and Tete Xiao and Spencer Whitehead and Alexander C. Berg and Wan-Yen Lo and Piotr Dollár and Ross B. Girshick},
  title = {Segment anything},
  booktitle = {{Proc. ICCV}},
  year = {2023}
}

@article{ravi2024sam2,
  author = {Nikhila Ravi and Valentin Gabeur and Yuan-Ting Hu and Ronghang Hu and Chaitanya Ryali and Tengyu Ma and Haitham Khedr and Roman R{\"a}dle and Chlo{\'e} Rolland and Laura Gustafson and Eric Mintun and Junting Pan and Kalyan Vasudev Alwala and Nicolas Carion and Chao-Yuan Wu and Ross Girshick and Piotr Doll{\'a}r and Christoph Feichtenhofer},
  title = {{SAM} 2: Segment Anything in Images and Videos},
  journal = {arXiv:2408.00714},
  year = {2024}
}

@InProceedings{Kim_2024_WACV,
    author    = {Kim, Jonghun and Park, Hyunjin},
    title = {Adaptive Latent Diffusion Model for {3D} Medical Image to Image Translation: Multi-Modal Magnetic Resonance Imaging Study},
    booktitle = {{Proceedings of the IEEE/CVF Winter Conference on Applications of Computer Vision (WACV)}},
    month     = {January},
    year      = {2024},
    pages     = {7604-7613}
}

@inproceedings{rombach2022high,
  title = {High-Resolution Image Synthesis with Latent Diffusion Models},
  author={Rombach, Robin and Blattmann, Andreas and Lorenz, Dominik and Esser, Patrick and Ommer, Bj{\"o}rn},
  booktitle = {{Proceedings of the IEEE/CVF Conference on Computer Vision and Pattern Recognition (CVPR)}},
  pages={10684--10695},
  year={2022}
}

@article{gu2023mamba,
  title = {Mamba: Linear-Time Sequence Modeling with Selective State Spaces},
  author={Gu, Albert and Dao, Tri},
  journal={arXiv preprint arXiv:2312.00752},
  year={2023}
}

@inproceedings{Liu2024VMamba,
  author = {Liu, Yue and Tian, Yunjie and Zhao, Yuzhong and Yu, Hongtian and Xie, Lingxi and Wang, Yaowei and Ye, Qixiang and Jiao, Jianbin and Liu, Yunfan},
  title = {{VMamba}: Visual State Space Model},
  booktitle = {{Advances in Neural Information Processing Systems (NeurIPS)}},
  year = {2024}
}

@article{xing2024i2imamba,
  author  = {Atli, Omer F. and Kabas, Bilal and Arslan, Fuat and
              Yurt, Mahmut and Dalmaz, Onat and {\c{C}}ukur, Tolga},
  title = {{{I2I-Mamba}}: Multi-modal medical image synthesis via
              selective state space modeling},
  journal = {arXiv preprint arXiv:2405.14022},
  year    = {2024}
}

@inproceedings{zhang2025structure,
  title = {Structure-Aware {MRI} Translation: Multi-modal Latent Diffusion Model with Arbitrary Missing Modalities},
  author={Zhang, Xinzhe and others},
  booktitle = {{International Conference on Medical Image Computing and Computer-Assisted Intervention}},
  year={2025},
  publisher={Springer Nature Switzerland},
  address={Cham},
  series={Lecture Notes in Computer Science}
}

@article{baid2021rsna,
  title = {The {RSNA}-{ASNR}-{MICCAI} BraTS 2021 benchmark on brain tumor segmentation and radiogenomic classification},
  author={Baid, Ujjwal and Ghodasara, Satyam and Mohan, Suyash and Bilello, Michel and Calabrese, Evan and Colak, Errol and Farahani, Keyvan and Kalpathy-Cramer, Jayashree and Kitamura, Felipe C and Pati, Sarthak and others},
  journal={arXiv preprint arXiv:2107.02314},
  year={2021}
}

@article{menze2015multimodal,
  title = {The multimodal brain tumor image segmentation benchmark ({BRATS})},
  author={Menze, Bjoern H and Jakab, Andras and Bauer, Stefan and Kalpathy-Cramer, Jayashree and Farahani, Keyvan and Kirby, Justin and Burren, Yuliya and Porz, Nicole and Slotboom, Johannes and Wiest, Roland and others},
  journal={IEEE Transactions on Medical Imaging},
  volume={34},
  number={10},
  pages={1993--2024},
  year={2015},
  publisher={IEEE}
}

@article{synthrad2023,
  title = {SynthRAD2023 Grand Challenge dataset: generating synthetic {CT} for radiotherapy},
  author={Thummerer, Adrian and van der Bijl, Erik and Galapon, Arturs and Verhoeff, Joost JC and Langendijk, Johannes A and Both, Stefan and van den Berg, Cornelis AT and Peter, Ronald},
  journal={arXiv preprint arXiv:2305.18331},
  year={2023}
}

@article{medsam2023,
  title = {Segment Anything in Medical Images},
  author  = {Ma, Jun and others},
  journal = {arXiv:2304.12306},
  year    = {2023}
}

\end{document}